\documentclass[showpacs,preprintnumbers,preprint]{revtex4}
\usepackage{amsmath}
\usepackage{graphicx}
\usepackage{dcolumn}
\usepackage{bm}

\newcommand{\beq}{\begin{equation}}
\newcommand{\eeq}{\end{equation}}
\newcommand{\eeg}{\end{gather}}
\newcommand{\beg}{\begin{gather}}
\newcommand{\bea}{\begin{array}}
\newcommand{\eea}{\end{array}}

\newcommand{\bz}{{\bf 0}}

\begin{document}

\title
{ New exactly solvable relativistic model with anomalous
interaction}
\author{Elena Ferraro}
\email{ferraro@fisica.unipa.it}
\affiliation{INFM, MIUR and\\
Dipartimento di Scienze Fisiche ed Astronomiche dell'Universit\`{a}
di Palermo \\Via Archirafi, 36 - I-90123 Palermo. Italy}
\author{ Antonino   Messina}
\email{messina@fisica.unipa.it}
 \affiliation{INFM, MIUR and\\
Dipartimento di Scienze Fisiche ed Astronomiche dell'Universit\`{a}
di Palermo \\Via Archirafi, 36 - I-90123 Palermo. Italy}
\author{A.G. Nikitin}
\email{nikitin@imath.kiev.ua}
\affiliation{Institute of Mathematics, National Academy of Sciences of Ukraine,\\
3 Tereshchenkivs'ka Street, Kyiv-4, Ukraine, 01601}
\date{\today}

\begin{abstract}
  A special class of Dirac-Pauli equations with
time-like vector potentials of external field is investigated. A new
exactly solvable relativistic model describing anomalous interaction
of a neutral Dirac fermion with a cylindrically symmetric external
e.m. field is presented. The related external field is a
superposition of the electric field generated by a charged infinite
filament and the magnetic field generated by a straight line
current. In non-relativistic approximation the considered model is
reduced to the integrable Pron'ko-Stroganov model.
\end{abstract}

\maketitle

\section{Introduction}

Exact solutions of relativistic wave equations are both very rare
and important. First they provide explicit solutions to concrete
physical problems free of inaccuracies  and  inconveniences  of
approximate methods. Secondly, such exact solutions can serve as
convenient basis sets for expanding the solutions of other physical
problems which are not necessarily exactly solvable.

A good survey of exact solutions of relativistic wave equations can
be found in \cite{bagrov}. Notwithstanding this book has been
published as long ago as in 1990, it continues to be a good
information resource on exactly solvable relativistic systems for
particles with spins 0 and 1/2. Surely this collection is not
exhausting: many new results have been obtained during two latest
decades, including the problems for the Dirac equation in lower
dimension space and the problems for neutral Dirac particles.

Exact solutions of the Dirac equation describing electrically
neutral particles with non-minimal interaction with an external
electromagnetic field are noteworthy. Physically, such solutions
have the very big application value since they can be used to model
the motion of a neutron in realistic situations. In particular they
have relations to the nuclear reactors security problems. Moreover,
magnetic trapping of neutrons is a subject of direct experimental
studies, refer, e.g., to \cite{trap}. Mathematically, the anomalous
interaction terms depending on tensor fields  dramatically reduce
the number of problems which can be solved using complete separation
of variables. In addition, just neutral particles anomalously
coupled to the external magnetic field give rise of the
Aharonov-Casher effect \cite{aharonov} with its interesting physical
and mathematical aspects.

The very possibility to solve a problem exactly stems from the
existence of a dynamical symmetry which is more extended than the
geometric symmetry of the problem. The famous examples of such
exactly solvable systems are the Kepler problems and isotropic
oscillator whose dynamical symmetries are defined by groups SO(4)
and U(3) respectively. One more well known example is the
Pron'ko-Stroganov (PS) problem \cite{pronko} which describes the
anomalous interaction of a non-relativistic electrically neutral
particle of spin 1/2 with the field of a straight line current. The
related dynamical symmetry of negative energy states is described by
group SO(3) while the geometrical symmetry of the system is reduced
to the rotation group in two dimensions, i.e., to SO(2). Let us
stress that the PS problem was formulated for the
Schr\"odinger-Pauli equation for neutral particles, i.e., it is
essentially nonrelativistic.

Paper \cite{pronko} was followed by a number of publications devoted
to exactly solvable problems for neutral particles. In particular,
the supersymmetric aspects of the PS model were investigated in
\cite{voronin}--\cite{planar}, more realistic models based on
magnetic field produced by current of thin filament were discussed
in \cite{blum} and \cite{Hau} following the non-relativistic
approach. A rather completed study of the Dirac-Pauli equation for
neutral particles can be found in \cite{shish}, the case of purely
electric time-independent external field was studied in
\cite{bruce}. However, an exactly solvable relativistic analogue of
the PS problem was not known till now.

We can add that searching for exact solutions of Dirac equation
belongs to evergreen problems, apparently the most recent result in
this field can be found in \cite{last}. Exactly solvable
two-particle Dirac equations are discussed, e.g.,  in \cite{Duv}.
For exact solutions of relativistic wave equations for particles
with higher and arbitrary spins see \cite{N1} and \cite{N2}.

In the present paper we discuss a certain class of relativistic
problems describing the anomalous interaction of the Dirac fermion
with an external electromagnetic field. The considered equations
admit an effective reduction to equations invariant with respect to
the 1+2 dimensional Galilei group, which can be made by using the
light cone coordinates.

Let us remind that the light cone coordinates was introduced by
Dirac \cite{Dirac} in an attempt to formulate a relativistic
dynamics with the direct interaction. Then it was recognized that
these coordinates present powerful tools for solution of
relativistic wave equations which include plane wave potentials or,
more generally, if the potentials do not depend on a scalar products
of the coordinate vector with a constant time-like vector
\cite{Redmond}, \cite{Bergou}.

 We show that the considered  class of equations includes new
exactly solvable relativistic systems and we study in detail one of
them, namely the one closely related to both the Pron'ko-Stroganov
problem and relativistic problems for the neutron interacting with
an external field. The corresponding external field is a
superposition of the magnetic field generated by the straight line
constant current and the electric field of a charged infinite
filament. In spite of that our main goal is to present an exactly
solvable problem for neutral fermions, for the sake of generality we
consider also more general problems with both the minimal and
anomalous interactions. We also indicate an exactly solvable model
of this general type, see Section 7.

\section{Dirac-Pauli equations and reduction SO(1,3) $\rightarrow$
HG(1,2)} Consider the Dirac-Pauli equation for a charged particle
which interacts anomalously with an external electromagnetic field:
\beq\label{1}\left(\gamma^\mu \pi_\mu-m- \lambda
S^{\mu\nu}F_{\mu\nu}\right)\psi=0.\eeq Here
$\pi_\mu=p_{\mu}-eA_\mu$, $p_{\mu}={\mathrm
i}\frac{\partial}{\partial x^\mu}, A_\mu$ are components of the
vector-potential of the external electromagnetic field, $\gamma^\mu$
are Dirac matrices satisfying the Clifford algebra
\beq\label{2}\gamma^\mu \gamma^\nu+\gamma^\nu
\gamma^\mu=2g^{\mu\nu}, \eeq $g^{\mu\nu}$ is the metric tensor whose
nonzero elements are $g^{00}=-g^{11}=-g^{22}=-g^{33}=1$,
$S^{\mu\nu}={\mathrm i}\left(\gamma^\mu\gamma^\nu-
\gamma^\nu\gamma^\mu\right)/4$ is the spin tensor,
$F_{\mu\nu}={\mathrm i}[\pi_\mu,\pi_\nu]$ is the tensor of the
electromagnetic field such that $F_{0a}=-E_a,
F_{ab}=\varepsilon_{abc}B_c$ where $E_a$ and $B_c$ are components of
vectors of the electric and magnetic field strengthes. In addition,
$e$ and $\lambda$ denote particle charge and constant of anomalous
coupling. The latest is usually represented as
\beq\label{lambda}\lambda=g\mu_0\eeq where
 $\mu_0$ is the Bohr magneton and $g$
is the Land\'e factor.

We use Heaviside units with $\hbar=c=1$.

Equation (\ref{1}) describes both the minimal and anomalous
interaction of the Dirac fermion with an external electromagnetic
field. Setting in (\ref{1}) $\lambda=0$ and supposing $e\neq0$ we
come to the equation describing the anomalous interaction only,
while for $e=0,\ \lambda\neq0$ we obtain the Dirac-Pauli equation
describing a neutral fermion. In the latest case the parameter $g$
in (\ref{lambda}) is just the contribution to the Land\'e factor
arising from the  anomalous magnetic moment.

Equation (\ref{1}) is transparently invariant with respect to the
Lorentz group SO(1,3) which transforms time and space variables
$x_0, x_1,x_2,x_3$ between themselves. Among the subgroups of this
group there is the homogeneous Galilei group HG(1,2) which includes
the transformations of variables $\tau,x_1,x_2$ where
$\tau=(x_0+x_3)/2$ (for all non-equivalent subgroups of SO(1,3) and
of the Poincar\'e group see \cite{Patera}).

To  search for exactly solvable problems based on equations
(\ref{1}) we restrict ourselves to special class of external fields,
which makes it possible to expand solutions of (\ref{1}) via
solutions of reduced equations invariant w.r.t. group HG(1,2). In
other words, we will discuss such external fields for which these
reduced equations be integrable.

To this end we first suppose that the vector-potential
${A=(A_0,A_1,A_2,A_3)}$ be light-like, i.e., \beq\label{light}A_\mu
A^\mu=0.\eeq This condition can be always satisfied up to gauge
transformations and so it does not lead to any loss of generality.
Then we restrict ourselves to the vector-potentials of the following
special form compatible with (\ref{light}):
\beq\label{ddd}A=(\varphi,0,0,\varphi)\eeq where $ \varphi$ is a
function of time and spatial variables. In addition we suppose that
$\varphi$ depends on three variables only, namely,
\beq\label{dddd}\varphi=\varphi(\tau,x_1, x_2).\eeq

Vector-potentials (\ref{ddd}), (\ref{dddd}) satisfy the Lorentz
gauge condition $p_\mu A^\mu=0$ identically and the invariants of
the related external field are both equal to zero, i.e.,
\beq\label{zero}{F_{\mu\nu}F^{\mu\nu}}=0
 \texttt{ and
} {\frac12
\varepsilon_{\mu\nu\rho\sigma}F^{\mu\nu}F^{\rho\sigma}}=0.\eeq

 For
convenience we fix a nonstandard realization of the Dirac matrices
and set
\beq\label{4}\bea{l}\gamma^0=\left(\bea{cc}\bz&I\\I&\bz\eea\right),\
\gamma^3= \left(\bea{rl}\bz&I\\-I&\bz\eea\right),\ \
\gamma^\alpha={\mathrm
i}\left(\bea{cc}\sigma_\alpha&\bz\\\bz&-\sigma_\alpha\eea\right)\eea\eeq
where $\alpha=1,2, \ \ \sigma_\alpha$ are Pauli matrices, $\bz$ and
$I$ are the $2\times2$ zero and unit matrix correspondingly. Then
\beq\label{5}S^{0\alpha}=\frac12
\left(\bea{cc}\bz&\sigma_\alpha\\
-\sigma_\alpha&\bz\eea\right), \ \ S^{3\alpha}=\frac12\left(\bea{cc}\bz&\sigma_\alpha\\
\sigma_\alpha&\bz\eea\right).\eeq

 System (\ref{1}) with a particular class of the vector-potential
given by equations (\ref{ddd}) and (\ref{dddd}) is homogeneous with
respect to the sum  of independent variables ${x_0+x_3}$. Thus it is
convenient to rewrite it in the light cone variables
\[\tau =\frac12(x_0- x_3) \texttt{ and  } \xi=\frac12(x_0+ x_3).\] As a
result we obtain: \beq\label{01} L\psi\equiv\left({\tilde
\gamma}_\mu {\tilde \pi}^\mu-m- \lambda \eta_\alpha
F_{\alpha}\right)\psi=0\eeq where $ F_{\alpha}=\frac{\partial
\varphi}{\partial x_\alpha}, \alpha=1,2,$\begin{gather}\nonumber
{\tilde \gamma}_0=\gamma_0+\gamma_3,\ {\tilde\gamma}_3=
\frac12(\gamma_0-\gamma_3), \ {\tilde \gamma}_\alpha=\gamma_\alpha,
\\\label{02} \eta_\alpha=\frac12(
{\gamma}_0{\gamma}_\alpha+{ \gamma}_3{ \gamma}_\alpha),\ {\tilde
\pi}_0={\mathrm i}\frac{\partial}{\partial \tau}-2e\varphi,\\
\nonumber
 {\tilde
\pi}_3=2P_{3}=2{\mathrm i}\frac{\partial}{\partial \xi},\ {\tilde
\pi}_\alpha=p_\alpha=-{\mathrm i}\frac{\partial}{\partial x_\alpha}
\,\ \end{gather} and summation w.r.t.
 repeated indices $ \mu$ and
 $ \alpha$ is imposed over the values $ {\mu=0,1,2,3}$ and $
{\alpha=1,2}$ respectively. In addition, we impose on solutions of
(\ref{01}) the standard condition of square integrability and ask
for $\psi\to0$ when $x_\alpha\to0$.

Operator $P_3$ commutes with $L$ and so is a constant of the motion
for equation (\ref{01}). Let us expand  solutions of this equation
via eigenvectors $\psi_M$ of $P_3$:
\beq\label{psiM}P_{3}\psi_M=M\psi_M\ \Rightarrow\
\psi_M=\exp({\mathrm i}M\xi)\psi(\tau,x_1,x_2).\eeq Let us denote
\beq\label{2c}\psi(\tau,x_1,x_2)=\left(\bea{c}
\rho(\tau,x_1,x_2)\\\chi(\tau,x_1,x_2)\eea\right)\eeq where $\rho$
and $\chi$ are two-component spinors.  Substituting (\ref{psiM}) and
(\ref{2c}) into (\ref{01})  and using realization (\ref{4}) of
$\gamma$-matrices
 we obtain the following system: \beq\label{50}\left({\mathrm i}\sigma_\alpha p_\alpha-m\right)\rho+
\left({\mathrm i}\frac{\partial}{\partial
\tau}-2e\varphi-\lambda{\sigma_\alpha F_\alpha} \right)\chi=0,\eeq
\beq\label{6}2M\rho-({\mathrm i}\sigma_\alpha p_\alpha+m)\chi=0.\eeq

It is easy to convince oneself from eqs.(\ref{50}), (\ref{6}) that,
without loss of generality we can assume that $M$ cannot take the
zero value. Indeed, setting $M=0$ in (\ref{6}) we reduce it to the
equation for $\chi$ which does not have non-trivial normalizable
solutions. Then, equating $\chi$ to zero in (\ref{50}), we obtain
the equation for $\rho$ whose normalizable solutions are trivial
also.

 It is interesting to note that this system is nothing
 but a (1+2)-dimensional version of the
Galilei-invariant L\'evi-Leblond equation \cite{ll1967} with
anomalous interaction, as can be immediately deduced by comparing
(\ref{50}) and (\ref{6}) with equation (52) for $e=k=0$ in
\cite{N0}. Solving eq.(\ref{6}) for $\rho$ under the condition \(M
\neq0\) and substituting it into eq.(\ref{50}) we obtain the
Schr\"odinger-Pauli equation for the two-component spinor $\chi$:
\beq\label{lala}{\mathrm
i}\frac{\partial\chi}{\partial\tau}=\left(\varepsilon_0+
 \frac{p^2}{2M}+2e\varphi
 +\lambda{\sigma_\alpha F_\alpha}\right)\chi\eeq where $\varepsilon_0=\frac{m^2}{2M} $ and
$p^2=p_1^2+p_2^2$.

Surely equations (\ref{lala}) are more easy to handle then the
initial equation (\ref{1}), since they include smaller numbers of
dependent and independent variables. In particular, a number of
exactly  (and quasi-exactly) solvable Scr\"odinger-Pauli equations
(\ref{lala}) is well studied, and many of them can be used to
construct solvable relativistic problems using the scheme inverse to
the previously proposed.

In Section 4 we use this idea to generate a relativistic analogue of
the PS problem.
\section{Cylindrically symmetric potentials}
Consider in more details a physically interesting subclass of
equations (\ref{1}), (\ref{ddd}), (\ref{dddd}) when the
corresponding potential $\varphi$ depends on the square
$x^2=x_1^2+x_2^2$ of 2-vector ${\bf x}=(x_1,x_2)$ and is independent
on  $\tau$. The related reduced equation (\ref{lala}) takes the form
\beq\label{lalala}{\mathrm
i}\frac{\partial\chi}{\partial\tau}=H\chi\eeq where
\beq\label{Ham}H=\varepsilon_0+
 \frac{p^2}{2M}+2e\varphi
 +\lambda\frac{\sigma_\alpha x_\alpha}{x}\frac{\partial \varphi}{\partial
x}.\eeq

Equation (\ref{lalala}) has three additional constants of motion,
namely, \beq\label{constants}\bea{l}  {P_0= {\mathrm
i}\frac{\partial}{\partial \tau}, \
 J_{12}=x_1p_2-x_2p_1+
\frac{{\mathrm i}}{2}\sigma_3}, \ \ Q=\sigma_1R_1 \eea\eeq where
$R_1$ is the reflection operator which acts on $\chi$ as
 follows: \[R_1\chi(\tau,x_1,x_2)=\chi(\tau,-x_1,x_2).\]

 Operators $P_0$ and $J_{12}$ are generator of shifts w.r.t. variable $\tau$
 and rotation generator correspondingly. They commute with the Hamiltonian
  (\ref{Ham}) and between themselves. Expanding solutions
of (\ref{lalala}) via complete sets of eigenfunctions of $P_0$ and
$J_{12}$ it is possible to separate variables in this equation.

Operator $Q$ represents a discrete symmetry w.r.t. the reflection of
the first coordinate axis. It commutes with $P_0$ and $H$
(\ref{Ham}) but {\it anticommutes} with $J_{12}$. It follows from
the above that eigenvalues of $P_0$ and $H$ should be degenerated
w.r.t. the sign of eigenvalues of $J_{12}$.

Notice that equation (\ref{lalala}) admits other discrete symmetries
like reflections of $x_2$ or both $x_1$ and $x_2$. But all such
additional symmetries are either rotation transformations or
products of reflection $Q$ and rotations.

Let us separate variables in equation (\ref{lalala}). First we
define the eigenvectors of $P_0$ which have the following form:
\beq\label{41}\chi_{\varepsilon }=\exp(-{\mathrm
i}\varepsilon\tau)\chi({\bf x}).\eeq Then, substituting
eq.(\ref{41}) into eq.(\ref{lalala}) we obtain the equation
\beq\label{ala}\varepsilon\chi=H\chi\eeq where $H$ is Hamiltonian
(\ref{Ham}).

In addition to the coupling constants $e$ and $\lambda$,
equation(\ref{ala}) includes two parameters, i.e., $\varepsilon$ and
$M$. For a fixed non-zero $M$
   this equation defines an
eigenvalue
 problem for $\varepsilon$.

 Now we can use the symmetry of (\ref{ala}) w.r.t. the rotation group
 (whose generator is $J_{12}$)
 to
 separate
radial and angular variables. To do this we rewrite equation
(\ref{ala}) in terms of angular variables, i.e., set
$x_1=x\texttt{cos}\theta, \ x_2=x\texttt{sin}\theta$,
$r=2M|{\tilde\lambda}|x$ (where $\tilde\lambda$ is a normalizing
parameter), and expand $\chi$ via eigenfunctions of the angular
momentum operator $J_{12}$: \beq\label{s1}\chi=  C_k \chi_k,\ \ \
\chi_k=\frac{1}{\sqrt{r}}
\left(\bea{c}\texttt{exp}(\texttt{i}(k-\frac12)\theta)\phi_1\\
\epsilon\texttt{exp}(\texttt{i}(k+\frac12)\theta)\phi_2\eea\right)\eeq
where $C_k$ are constants, $\epsilon=\tilde\lambda/|\tilde\lambda|,\
\ \phi_1$ and $\phi_2$ are functions of $r$ and summation is imposed
over the repeated indices
$k=\pm\frac12,\pm\frac32,\pm\frac52\cdots$.

In the following we restrict ourselves to solutions $\chi_k$ which
correspond to non-negative values of $k$. Then solutions with $k$
negative would be obtained by acting on $\chi_k$ by operator $Q$
(\ref{constants}).

 Substituting eq.(\ref{s1}) into eq.(\ref{ala}) we come to the following system:
\beq\label{s2}\bea{l}H_k\phi\equiv\left(-\frac{\partial^2}{\partial
r^2}+k(k-\sigma_3)
\frac{1}{r^2}+2e\varphi+\sigma_1\frac{\lambda}{\tilde\lambda}
\frac{\partial\varphi}
{\partial{r}}\right)\phi={\tilde\varepsilon}\phi\eea\eeq  with
$\phi=\texttt{column}(\phi_1,\ \phi_2)$, and \beq\label{va} \tilde
\varepsilon=(\varepsilon-\varepsilon_0)/2M{\tilde \lambda}^2.\eeq

Thus we reduce (\ref{ala}) to the system of two ordinary
differential equations for radial functions, given by formula
(\ref{s2}). Its solutions must be normalizable and vanish at $r=0$.
For some types of potential $\varphi$ (and particular restrictions
imposed on the coupling constants $e$ and $\lambda$) this system is
integrable and its solutions can be expressed via special functions.
In the following section we consider an example of integrable
equation (\ref{s2}) which corresponds to a neutral particle
interacting anomalously with an external field.

 \section{Relativistic  analog of PS problem}

Let us set $e=0$ in (\ref{50}), (\ref{6}) and choose the following
particular realization for the potential $\varphi$:
\beq\label{pot}\varphi=\omega\log(x)\eeq where $\omega$ is a
constant.  Then the related equations (\ref{lalala}) and (\ref{s2})
are reduced to the following forms
\beq\label{03}{\varepsilon}'\chi=\left(\frac{p^2}{2M}+\tilde\lambda\frac{\sigma_\alpha
x_\alpha} {x^2}\right)\chi,\ \
{\varepsilon}'={\varepsilon}-{\varepsilon}_0\eeq and
\beq\label{s22}H_k\phi\equiv\left(-\frac{\partial^2}{\partial
r^2}+k(k-\sigma_3) \frac{1}{r^2}+\sigma_1\frac{1}
{{r}}\right)\phi={\tilde\varepsilon}\phi\eeq correspondingly,
provided we set $\tilde\lambda=\omega\lambda$.

The electromagnetic field whose potential is defined by relations
(\ref{ddd})
 and (\ref{pot})
  has a transparent physical meaning. Namely, it is a superposition of the
electric field ${\bf E}=(E_1,E_2,E_3)$ whose components are
\beq\label{E}E_1=\omega\frac{x_1}{x^2},\ \
E_2=\omega\frac{x_2}{x^2},\ \  E_3=0\eeq and the magnetic field
${\bf B}= (B_1,B_2,B_2)$ with
\beq\label{B}B_1=-\omega\frac{x_2}{x^2},\ \
B_2=\omega\frac{x_1}{x^2},\ \ B_3=0.\eeq Such an electric field can
be identified as the field of a charged infinite filament coinciding
with the third coordinate axis. Let us designate the charge density
of this filament by $\rho$ then the coupling constant $\omega$
should be equal to $2\rho$. On the other hand, the magnetic field
${\bf B}$ is nothing but the field of a straight line constant
current $j$ directed along the third coordinate axis
 provided the coupling constant $\omega$ be equal to
 $2j$. Of course the related charge density and current should be
 equal between themselves, i.e., \beq\label{jw}j=\rho=\omega/2.\eeq

Let us show that equation (\ref{03}) is exactly solvable and find
its solutions. The simplest way to prove integrability of (\ref{03})
is to make the unitary transformation \beq\label{08}\bea{l}\chi\to
\chi'=U\chi,\
{\varepsilon'}-\left(\frac{p^2}{2M}+\tilde\lambda\frac{\sigma_\alpha
x_\alpha} {x^2}\right)\to U\left(
{\varepsilon'}-\left(\frac{p^2}{2M}+\tilde\lambda\frac{\sigma_\alpha
x_\alpha} {x^2}\right)\right)U^\dag\eea\eeq where
$U=\frac{1}{\sqrt2}(1-{\mathrm i}\sigma_3).$ As a result we reduce
(\ref{03}) to the following form:
\beq\label{09}{\varepsilon}'\chi=\left(\frac{p^2}{2M}-
2\tilde\lambda\frac{S_1 x_2-S_2 x_1} {x^2}\right)\chi\eeq where
$S_1=\frac12\sigma_1$ and $S_2=\frac12\sigma_2$ are spin matrices
and in accordance with eqs.(\ref{lambda}), (\ref{pot}) and
(\ref{jw}) $\tilde\lambda=\lambda\omega=2g\mu_0j$.

 For a fixed $M$ and up to the value of the coupling constant equation (\ref{09}) coincides
with the Schr\"odinger equation for a neutral particle minimally
interacting with the field generated by an infinite thin current
filament (in our case the standard coupling constant is multiplied
by factor 2). This equation was studied in numerous papers starting
with \cite{pronko} and continuing with \cite{voronin}--\cite{planar}
and many others. It has the following nice properties:
\begin{itemize}
\item  equation  (\ref{09}) admits a hidden dynamical symmetry with respect
to group $SO(3)$ for negative eigenvalues $\tilde\varepsilon$, group
$SO(1,2)$ for $\tilde\varepsilon$ positive  and group $E(2)$ for
${\tilde\varepsilon}=0$ \cite{pronko};
\item it possesses a hidden supersymmetry \cite{voronin};
\item using any of the above mentioned properties the equation can be integrated in closed form
 \cite{pronko}-\cite{Hau}.
\end{itemize}

Since equation (\ref{03}) is unitary equivalent to eq.(\ref{09}) it
succeeds the above mentioned properties. In particular, eigenvalues
${\varepsilon}'$ are the same in both equations (\ref{03}) and
(\ref{09}).

\section{Relativistic and quasi relativistic energy levels}

In the next section we will present exact solutions of equation
(\ref{03})  for coupled states and define the related eigenvalues
${\varepsilon}'$. In fact these eigenvalues are well known, and
using directly results of paper \cite{pronko} (or of the papers
\cite{voronin}-\cite{blum}) we can immediately write
${\varepsilon}'$ in the following form:
\beq\label{011}\varepsilon'=-\frac{2\tilde\lambda^2M}{N^2}\eeq where
$N$ is a positive natural number.

Eigenvalues (\ref{011}) are degenerated since they do not depend on
eigenvalues $k$ of the angular momentum operator $J_{12}$. The
degeneration factor is equal to $2k+1$, and the quantum number $N$
can be represented as \beq\label{N}N=2(n+k)+1\eeq where $n$ is a
natural number \cite{pronko}, \cite{Hau}.

 Using (\ref{011}) we already can find energy levels
for the initial relativistic problem. Indeed, since ${P}_0=p_0+p_3$
and ${P_3}=\frac12(p_0-p_3)$, it is possible to write analogous
relations for eigenvalues $E$ of ${ p}_0,\ \kappa$ of ${ p}_3$ and
$\varepsilon, M$: \beq\label{001}\varepsilon=E+\kappa,\ \
2M=E-\kappa.\eeq Then, using definitions (\ref{03})  and (\ref{001})
for $\varepsilon'$, $E$ and $M$ we find from (\ref{011}) the
relativistic energy spectrum:
\beq\label{012}E=\frac{m}{K+\tilde\kappa}+\kappa\eeq where
\beq\label{me}
K=\sqrt{1+\tilde\kappa^2+\frac{\tilde\lambda^2}{N^2}},\ \
\tilde\kappa=\frac{\kappa}{m},\ \ {\tilde\lambda}=2\mu_0gj.\eeq

We see that in spite of the fact that the neutron motion along the
third coordinate axis is free, the third component of momentum
$\kappa$ makes a rather non--trivial contribution into the values of
energy levels (\ref{012}). In accordance with (\ref{012})
$E>\kappa$, and so the condition $M\neq0$  is actually satisfied.

The most simple expression for energy levels corresponds to the
particular value $\kappa=0$:
\beq\label{0131}E=\frac{m}{\sqrt{1+\frac{{\tilde\lambda}^2}{N^2}}}
 \eeq which
for small $\tilde\lambda$ becomes \begin{gather}
E=m\left(1-\frac{\tilde\lambda^2}{2N^2}\right)+\cdots=
m-\frac{2m(g\mu_0j)^2}{N^2}+\cdots.\label{013}\end{gather}

Up to the rest energy term $m$ the approximate energy levels
(\ref{013}) are exactly the same as in the non-relativistic PS
problem \cite{pronko}, \cite{Hau}--\cite{blum}. In particular both
the approximate and exact levels given by equations (\ref{012}),
(\ref{0131}) and (\ref{013}) are degenerated with respect to
eigenvalues $k$ of the third component of angular momentum which is
a constant of motion for the considered system. Like in
\cite{pronko} this degeneration is caused by a hidden dynamical
symmetry of the system.

Let both $\tilde\lambda$ and $\tilde\kappa$ are small. Expanding $E$
(\ref{013}) in power series of $\tilde\lambda$ and $\tilde\kappa$ we
obtain the quasi relativistic approximation for the energy levels:
\beq\label{qr}E\approx m+\frac{\kappa^2}{2m}-\frac{\kappa^4}{8m^3}-
\frac{m\tilde\lambda_\kappa^2}{2N^2}\eeq where
\beq\label{kappa}\tilde\lambda_\kappa=(1-\tilde\kappa)\tilde\lambda.\eeq

The first tree terms in (\ref{qr}) represent respectively the rest
energy, the kinetic energy of the free motion along the third
coordinate axis and the relativistic correction to this energy. The
last (dynamical) term in (\ref{qr}) is quite similar to the
corresponding non-relativistic term (compare with (\ref{013})), but
includes the corrected coupling constant $\tilde\lambda_\kappa$
instead of $\tilde\lambda$.

Consider also the ultrarelativistic situation when large
$\tilde\kappa$ is large but $\tilde\lambda$ is still small. Then the
energy values (\ref{013}) can be expanded as
\beq\label{Nu}E=\sqrt{m^2+\kappa^2}
-m\delta\frac{\tilde\lambda^2}{2N^2}+\cdots\eeq where the dots
denote the terms of order $\tilde\lambda^4$, and
\beq\label{muk}\delta=\left(2\sqrt{1+\tilde\kappa^2}-2\tilde\kappa-\frac1
{\sqrt{1+\tilde\kappa^2}}\right).\eeq  Comparing (\ref{Nu}) with
(\ref{013}) we recognize that the relativistic binding energy levels
include the additional multiplier $\delta$ which considerable
differs from 1 for ultrarelativistic $k$.

Notice that since the electromagnetic field defined by relations
(\ref{ddd}) and (\ref{dddd}) has no components in $x_3$ direction,
the motion of the particle in this direction is free. This motion
can be quantized by imposing the periodic boundary condition. Then
\beq\label{p3}\kappa=\frac{2\pi \tilde N}{L},\ \tilde N=0, \pm1,
\pm2,\cdots \eeq and energy levels (\ref{012})--(\ref{Nu}) be
labeled by the pairs of quantum numbers $N$ and $\tilde N$.

\section{Exact solutions for bound states}

To find the solutions of equation (\ref{03}) we use the fact that
the Hamiltonian \beq H_k=\left(-\frac{\partial^2}{\partial
r^2}+k(k-\sigma_3) \frac{1}{r^2}+\sigma_1\frac{1} {{r}}\right)\eeq
can be factorized as \beq\label{s3}H_k=a_k^+a_k+C_k\eeq where
\[a_k=\frac{\partial}{\partial r}+W_k,\  \ a_k^+=-
\frac{\partial}{\partial r}+W_k,\ \ C_k=-\frac{1}{(2k+1)^2}\] and
$W$ is a {\it matrix superpotential}
\beq\label{s4}W_k=\frac{1}{2r}\sigma_3-\frac{1}{2k+1}\sigma_1-
\frac{\left(k+\frac12\right)}{r}.\eeq

It can be verified by direct calculation that
\[H_k^+=a_ka_k^++C_k= -\frac{\partial^2}{\partial r^2}+(k+1)(k+1-\sigma_3)
\frac{1}{r^2}+\sigma_1\frac1r\] i.e.,  the superpartner Hamiltonian
$H_k^+$ for $H_k$ is equal to $H_{k+1}$. Thus the eigenvalue problem
(\ref{s2}) possesses a supersymmetry with shape invariance and so it
can be solved using the standard technique of the supersymmetric
quantum mechanics \cite{gen}. We will not reproduce the related
routine calculations whose details can be found in \cite{Hau},
\cite{planar}  but restrict ourselves to the presentation of the
solutions of eq.(\ref{03}).

The ground state solutions $\phi(0,k;r)=\texttt{column
}(\phi_1(0,k;r),\phi_2(0,k;r))$ are square integrable and
normalizable solutions of equation $a_k{\phi(0,k;r)}=0$. They can be
expressed in the following form:
\begin{gather}\phi_1(0,k;r)= r^{k+1}K_1\left(\frac{r}{2k+1}\right),\ \ \label{0}
\phi_2(0,k;r)= -r^{k+1}K_0\left(\frac{r}{2k+1}\right)\end{gather}
where $K_0$ and $K_1$ are the modified Bessel functions. The
corresponding eigenvalue $\tilde\varepsilon_k$ in (\ref{03}) and
(\ref{s2}) is equal to $-\frac{1}{(2k+1)^2}$.

Solutions corresponding to the first excited state, i.e., to $n=1$
are $\phi(1,k;r)=a_k^+\phi(0,k+1;r)$, or, being written
componentwise,
\begin{widetext}\begin{gather}\label{11}\begin{split}\phi_1(1,k;r)&
=-\left(\frac{\partial }{\partial
r}+\frac{k}{r}\right)\phi_1(0,k+1;r)
-\frac{1}{(2k+1)}\phi_2(0,k+1;r)\\&{}=
\frac{4(k+1)}{(2k+1)(2k+3)}r^{k+2}K_0\left(\frac{r}{2k+3}\right)-(2k+1)r^{k+1}K_1\left(\frac{r}{2k+3}\right),
\\\phi_2(1,k;r)=&{}-\left(\frac{\partial }{\partial r}
+\frac{k+1}{r}\right)\phi_2(0,k+1;r)-\frac{1}{(2k+1)}\phi_1(0,k+1;r)\\&{}=
(2k+3)r^{k+1}K_0\left(\frac{r}{2k+3}\right)
-\frac{4(k+1)}{(2k+1)(2k+3)}r^{k+2}K_1\left(\frac{r}{2k+3}\right).\end{split}\end{gather}\end{widetext}
The corresponding eigenvalue $\tilde\varepsilon_k$ is equal to
$-\frac{1}{(2(k+1)+1)^2}=-\frac{1}{(2k+3)^2}$.

Finally, solutions which correspond to an arbitrary value of the
quantum number $n>0$ can be represented as
\begin{gather*}\phi(n,k;r)=a_k^+a_{k+1}^+\cdots a_{k+n-1}^+\phi(0,k+n;r),\ \
n=1,2,\cdots
\end{gather*} which gives rise to the recurrence
relations
\begin{gather}
\begin{split}
\phi_1(n,k;r)=& -\frac{\partial }{\partial r}\phi_1(n-1,k+1;r)
-\frac{k}{r}\phi_1(n-1,k+1;r)\\
&{}+\frac{1}{2k+1}\phi_2(n-1,k+1;r),\\
\phi_2(n,k;r)=& -\frac{\partial }{\partial r}\phi_2(n-1,k+1;r)
-\frac{k+1}{r}\phi_2(n-1,k+1;r)\\
&{} +\frac{1}{2k+1}\phi_1(n-1,k+1;r).
\end{split}
\label{n}
\end{gather}
 The related eigenvalue $\tilde\varepsilon_k$ is
given by relations (\ref{011}) and (\ref{N}).

It is now possible to present exact solutions of the initial
Dirac-Pauli equation defined by relations (\ref{1}), (\ref{ddd}) and
(\ref{pot}). In accordance with the above such solutions are labeled
by the main quantum number $N$ which can be expressed by equation
(\ref{N}), and by eigenvalues $\kappa$ and $k$ of the third
component of momenta
 and total orbital momentum. Using equations
(\ref{psiM}), (\ref{2c}), (\ref{6}), (\ref{41}), (\ref{s1}) and
(\ref{0})--(\ref{n}) we find these solutions in the following form:
\begin{equation}\label{Sol}\bea{l}\psi_{n,\kappa, k}=\frac{1}{\sqrt{2\pi
Lr}}\exp(-{\mathrm i}(E x_0-\kappa x_3))
\left(\bea{c}{\exp}(i(k-\frac12)\theta)\eta_1(n,k;r)\\
{\exp}(i(k+\frac12)\theta)\eta_2(n,k;r)\\{\exp}(i(k-\frac12)
\theta)\phi_1(n,k;r)\\{\exp}(i(k+\frac12)\theta)\epsilon\phi_2(n,k;r)
\eea\right).\eea\eeq Here $k$ and $n$ are nonnegative natural
numbers,
\begin{gather}\label{Nik}r=\frac{\sqrt{x_1^2+x_2^2}}{r_0}, \ \
r_0=\frac{1}{M|\tilde\lambda|}=\frac{K+\tilde\kappa}{m|\tilde\lambda|
},\ \ \epsilon=\frac{\tilde\lambda}{|\tilde\lambda|}\end{gather}
where $\tilde\kappa=\kappa/m$, $E$, $K$ and $\kappa$ are given by
equations (\ref{012}), (\ref{me}) and (\ref{p3}),  $\phi_1(n,k;r)$
and $\phi_2(n,k;r)$ are functions defined by recurrence relations
(\ref{0}), (\ref{n}),\begin{gather}\begin{split}&\eta_1(n,k;r)=
\tilde\lambda\left(\frac{\partial}{\partial
r}+\frac{k}{r}\right)\epsilon\phi_2(n,k;r)+(K+\tilde\kappa)\phi_1(n,k;r),
\\&\eta_2(n,k;r)=\tilde\lambda\left(\frac{\partial}{\partial
r}-\frac{k}{r}\right)\phi_1(n,k;r)+(K+\tilde\kappa)\epsilon\phi_2(n,k;r).\end{split}\label{O}\end{gather}

Solutions (\ref{Sol}) are normalizable and tend to zero with
${r}\to0$. They are defined for non-negative eigenvalues $k$ of the
total angular momentum while solutions for $k$ negative can be
obtained acting on (\ref{Sol}) by the reflection operator $\hat
Q=i\gamma_0\gamma_2\gamma_3R_1$ where $R$ is the reflection of the
first spatial variable, i.e.,
$R_1\psi(x_0,x_1,x_2,x_3)=\psi(x_0,-x_1,x_2,x_3)$ and
$R_1\psi(x_0,r,\theta,x_3)=\psi(x_0,,r,-\theta,x_3)$. Using the
Dirac matrices (\ref{4}) we find the solutions with negative values
of $k$ in the following form:
\beq\label{solol}\bea{l}\psi_{n,\kappa,
k}=\frac{\texttt{i}}{\sqrt{2\pi Lr}}\exp({\mathrm i}(E x_0-\kappa
x_3))
\left(\bea{c}-\texttt{exp}(-i(k+\frac12)\theta)\eta_2(n,k;r)\\
-\texttt{exp}(i(\frac12-k)\theta)\eta_1(n,k;r)\\\epsilon\texttt{exp}(-i(k+\frac12)
\theta)\phi_2(n,k;r)\\-\texttt{exp}(i(\frac12-k)\theta)\phi_1(n,k;r)
\eea\right). \eea\eeq

Let us present explicitly the components of solutions (\ref{Sol}),
(\ref{solol}) for $n=0$ and $n=1$.
 If $n=0$ then the related functions $\phi_1(n,k;r)=\phi_1(0,k;r)$ and
$\phi_2(0,k;r)$ are given by equation (\ref{0}) while
$\eta_1(0,k;r)$ and $\eta_2(0,k;r)$ have the following form:
\begin{gather}\label{N00}\begin{split}&\eta_1(0,k;r)=
\Lambda_k\phi_1(0,k;r)
+\frac{\tilde\lambda(2k+1)}{r}\phi_2(0,k;r),\\&{}
\eta_2(0,k;r)=\Lambda_k\phi_2(0,k;r)\end{split}\end{gather} where
$\Lambda_k=\left(\frac{ \tilde\lambda}{2k+1}+K+\tilde\kappa\right).$
If $n=1$ the corresponding functions $\phi_1(1,k;r)$ and
$\phi_2(1,k;r)$ are given by equations (\ref{11}), and
\begin{gather}\label{AGN}\begin{split}\eta_1(1,k;r)&=\Lambda_{k+1}\phi_1(1,k;r)+\frac{\tilde\lambda(2k+1)}{r}
\phi_2(1,k;r)-\frac{2\tilde\lambda}{(2k+3)r}\phi_1(0,k+1;r),
\\\eta_2(1,k;r)&{}=\Lambda_{k+1}\phi_2(1,k;r)+\frac{2\tilde\lambda}{r}
\phi_1(1,k;r)+\frac{2\tilde\lambda}{(2k+3)r}\phi_2(0,k+1;r)\\&{}+
\frac{2(2k+1)\tilde\lambda}{r^2}\phi_1(0,k+1;r)\end{split}\end{gather}
where $\Lambda_{k+1}=\left(\frac{ \tilde\lambda}
{2k+3}+K+\tilde\kappa\right).$

Functions $\eta_a (a=1,2)$ in (\ref{N00}) and (\ref{AGN}) include
the terms proportional to $\phi_a$ (the first terms in the r.h.s.).
The remaining terms are small in comparison with $\phi_a$, which
results in similarity of probability distributions for neutrons in
our model to the distributions in the PS model, see Appendix.

\section{Discussion}

In Sections 2 and 3 we study a class of Dirac-Pauli systems  which
can be effectively reduced to a set of Schr\"odinger-Pauli
equations. The main inspiration for our research was to find an
integrable relativistic formulation of the non-relativistic PS
problem \cite{pronko}. This goal cannot be achieved by a
straightforward relativization of the PS problem since the
Dirac-Pauli equation for a neutral particle interacting with the
magnetic field generated by a filament current is not integrable.

In the present paper we succeed in obtaining an integrable
relativistic model which in many aspects can be treated as an
analogue of the PS model. To this end we introduce a superposition
of magnetic and electric fields which is not equivalent to the field
of straight current. Nevertheless, in the non-relativistic limit our
model is reduced to the PS one.  To justify this statement we return
to equation (\ref{jw}) and note that in the CGS units it takes the
following form: \beq\label{d1}\rho={j}/{c}\eeq where $c$ denotes the
velocity of light.

In accordance with (\ref{d1}) the required charge density is small
and tends to zero in the non-relativistic approximation when
$c\to\infty$. Thus in the non-relativistic limit the external field
which we consider reduces to the field used in the PS model. In
addition, the energy spectrum (\ref{012}) is reduced to the PS form,
see equation (\ref{013}). That is why we claim that the
non-relativistic limit of the model defined by equations (\ref{1}),
(\ref{ddd}) and (\ref{pot}) is exactly the PS model. This property
can be directly proved using the Foldy-Wouthuysen transformation
\cite{FW}.

Let us discuss the obtained energy levels for coupled states in the
quasi-relativistic approximation (\ref{qr}). Albeit the motion along
the third Cartesian coordinate is free,  the third momentum
component $\kappa$ makes a contribution into the effective coupling
constant $\tilde\lambda_\kappa$.  The origin of this contribution is
the anomalous interaction of neutron moving along the charged line
with the magnetic field generated by this line. In the rest frame
this motion is effectively changed by the current which flows in the
line in the opposite direction, which is in perfect accordance with
equation (\ref{kappa}).

The contribution of $\kappa$ into the effective coupling constant
$\tilde\lambda_\kappa$ (\ref{kappa}) is small. Namely, it is
proportional to the  inverse speed of light. However, it affects the
energy levels (\ref{qr}) much more than the relativistic correction
to the kinetic energy $-\kappa^4/8m^3$ which is proportional to the
squared inverse speed of light.

In the case of ultrarelativistic neutron motion along the charge and
current carrying lines the contribution of the related momentum into
the  coupling energy becomes very essential. As it follows from
(\ref{Nu}) the distances between the energy levels can significantly
differ from the non-relativistic ones since the multiplier $\delta$
(\ref{muk}) changes continuously from 0.121 (for
$\tilde\kappa\to-1$) to 4.121 (for $\tilde\kappa\to 1$).

 In conclusion, we have presented a new exactly solvable
problem for the Dirac-Pauli equation describing a neutral particle
which interacts anomalously with a rather particular external field
given by equations (\ref{E})-(\ref{jw}) having however a clear
physical meaning. This type of anomalous interaction is the key to
expand solutions of the problem via solutions of the
(1+2)-dimensional Levi-Leblond equation invariant with respect to
Galilei group. Moreover, the considered problem possesses a hidden
symmetry and supersymmetry which cause the (2n+1)-fold degeneration
of the energy levels given by equations (\ref{012}) and (\ref{N}).

 A natural question arises whether the considered relativistic
problem with its symmetries is unique or there are other problems
which can be effectively solved using reduction SO(1,3)$\to$HG(1,2).
In Sections 2 and 3 we study a certain class of such problems which
can be effectively reduced to radial equation (\ref{s2}) which is
exactly solvable when $e=0$ and $\varphi=\omega\log(x)$. We believe
that there are other exactly solvable equations (\ref{s2}) and at
least two of them can be immediately written down if we set
$\varphi=\alpha/x$ and consider the alternative cases $e=0$ and
$e\neq 0$. The related Dirac-Pauli equations (\ref{1}), (\ref{ddd}),
(\ref{dddd}) can be solved explicitly. We plane to study these and
probably other integrable models in future.

\vspace{5mm}

{\bf Acknowledgments}

\vspace{3mm}

 One of co-authors (A.G.N.)
acknowledges partial support by Miur project Cooperazione
Internazionale AF 2007 with National Academy of Sciences OD Ukraine
and Kwa-Zulu University Durban South Africa.

\appendix*
\section{Coupling constants and probability distributions}

 In the main text we
consider an idealized model with the infinite thin current filament
and charged line. To give an idea about its physical realizability
let us discuss the probability density which corresponds to found
solutions (\ref{Sol}).

We formulated our problem for neutrons. However, the obtained
results can be extended to other neutral particles which have
non-trivial magnetic moments. As an example we consider here the
sodium atoms in the ground state.

First we present in more transparent form the coupling constant
$\tilde\lambda$ and scaling interval $r_0$ (\ref{Nik}). Going from
the Heaviside units to CGS ones we should make the following changes
in equations (\ref{012}) and (\ref{Nik}):
\begin{gather}\label{NINNI} m\to mc^2,\ \ \kappa\to c\kappa,\
\tilde\kappa\to\tilde\kappa'=\frac{\kappa}{mc},\
\tilde\lambda\to\frac{\tilde\lambda}{\hbar c}=\tilde\lambda',\
r_0\to
\frac{2C_n}{|\tilde\lambda'|}\left(K+\frac{\kappa}{mc}\right)\end{gather}
 where $c$ is the velocity of light and $C_n$ is the Compton wave length
 for the neutron. The dimensionless constant $\tilde\lambda'$ and the $r_0$
can be represented as
\begin{gather}\label{AG}\tilde\lambda'=-\frac{g\alpha C_nN_c\hat j
}c=7.633\cdot10^{-7}\hat j,\ \ r_0=\frac{34.5\AA}{\hat
j}\left(K+\frac{\kappa}{mc}\right)\end{gather} where ${\hat
j}=j/\texttt{A}$ is the current measured in Amperes,
$\alpha=\frac{e^2}{\hbar c}=\frac1{137}$ is the fine structure
constant, $g=-3.82$ is the neutron Land\'e factor,
$N_c=C/e=6.242\cdot 10^{18}$ is the charge equal to one coulomb
measured in elementary charges. Surely for realistic current values
parameter $\tilde\lambda$ is small thus the expansions in power
series of $\tilde\lambda$ made in Section V was well grounded.

We formulated our problem for neutrons. However, the obtained
results can be extended to other neutral particles which have
non-trivial magnetic moments. As an example we consider here the
sodium atoms in the ground state. Then $|g|\to5.4, m\to 23m$, and
the parameters (\ref{AG}) are transformed  to the following ones:
\begin{gather}\label{AGN}\tilde\lambda'=4.68\times 10^{-8}\hat j,\ \
r_0=\frac{24.42\AA}{\hat
j}\left(K+\frac{\kappa}{mc}\right).\end{gather}

Consider now solutions (\ref{Sol}) and evaluate the corresponding
probability density:
\begin{gather}\label{Nik12}\begin{split} & \bar\psi_{n,\kappa,k}
\gamma_0r\psi_{n,\kappa,k}= C^2_{n,\kappa,k}\left(\phi_1^2+\phi_2^2+
\tilde\lambda'\left(K+\kappa'\right)\epsilon\delta\frac{\partial(\phi_1\phi_2)}{\partial
r}+\frac{\delta\tilde\lambda'^2}{r}\frac{\partial(\phi_2^2-\phi_1^2)}{\partial
r}\right)\end{split}
\end{gather}
 where $C_{n,\kappa,k}$ is a normalization constant, $\delta=
 \frac1{\left(K+\kappa'\right)^2+1}$.
 In particular, for $n=0$,
\begin{gather}\label{LL}\bar\psi_{0,\kappa,k}\gamma_0r\psi_{0,\kappa,k}=
C^2_k\left(\phi_1^2+\phi_2^2- \epsilon\delta_1
\frac{\phi_1\phi_2}{r}+\delta_2\frac{\phi_2^2}{r^2}\right)\end{gather}
where $\phi_1$ and $ \phi_2$ are functions defined in (\ref{0}) and
\begin{gather}\label{delta}\delta_1=2\tilde\lambda'\left((2k+1)
\left(K+\kappa'\right)+\tilde\lambda'\right)\delta
 ,\ \ \delta_2=
\tilde\lambda'^2(2k+1)^2\delta.\end{gather} Analysing (\ref{LL}) we
conclude that the last two terms in the brackets are small. First
they include the small multiplier $\tilde\lambda'$  (\ref{AG}).
Secondly, for $k>1/2$ functions $\frac{\phi_1\phi_2}{r}$ and
$\frac{\phi_2^2}{r^2}$ are negligibly small in comparison with
$\phi_1^2+\phi_2^2$. The same statement is correct for equation
(\ref{Nik12}) which can be proven with using the identities
\begin{gather*}\frac{\partial K_0(\lambda r)}{\partial r}=-
\lambda K_1(\lambda r),\ \ \frac{\partial K_1(\lambda r)}{\partial
r}=-\lambda K_0(\lambda r)-\frac1r K_1(\lambda r).
\end{gather*}

Thus, practically without loss of accuracy, we can write
\begin{gather}\label{mik}\bar\psi_{n,\kappa,k}
\gamma_0r\psi_{n,\kappa,k}\approx
C^2_{n,\kappa,k}\left(\phi_1^2+\phi_2^2\right)\end{gather} and so
the probability distribution calculated for our relativistic problem
 is virtually the same as the one obtained in \cite{blum} and
\cite{Hau} for the non-relativistic PS problem.

 Thus the main statements presented in \cite{blum} and \cite{Hau}
concerning the possibility in principle to observe experimentally
the neutrons and sodium atoms trapped by the current filament, can
be generalized to our model, if we restrict ourselves to small
$\tilde\kappa$.
 We will not repeat the reasonings given in
the mentioned papers but remind that reasonable current values are
$j \approx 50mA$ for trapping neutrons and $j\approx 400 mA$ for
trapping the sodium atoms in the ground state.

The principally new feature of the relativistic model is the
essential dependence of the coupling energy and of the scaling
parameter $r_0$ on the third component of momenta. In accordance
with  (\ref{Nu}), (\ref{muk}) and (\ref{AG}), (\ref{AGN}) for
ultrarelativistic values of $\tilde\kappa\to1$ the coupling energy
obtains a non-unit multiplier which can  increase up to $\sqrt{2}+1$
and $r_0$ has a multiplies which can increase up to 4 comparison
with the case when $\tilde\kappa=0$ when these multipliers are the
unit one. These facts can open new ways for experimental search for
trapped neutral particles with non-trivial magnetic moments.

\end{document}